# Shear Stress Build-up Under Constant Strain Conditions in Soft Glassy Materials

Vivek Kumar[1], Gareth H McKinley[2], and Yogesh M Joshi[1,*]

[1]Department of Chemical Engineering, Indian Institute of Technology Kanpur, Kanpur, Uttar Pradesh, India, 208016.

[2]Hatsopoulos Microfluids Laboratory, Department of Mechanical Engineering, Massachusetts Institute of Technology, Cambridge, Massachusetts, USA, 02139

[*]Corresponding Author, email: joshi@iitk.ac.in

## Abstract

In this work, we investigate the transient rheological behavior of two soft glassy materials: a clay dispersion and a silica gel, emphasizing their unconventional shear stress build-up behavior under conditions of constant imposed strain. For both materials, the elastic modulus and static yield stress undergo time-dependent evolution or aging. In addition, following an intense period of pre-shearing (i.e. shear-melting or destructuration), the material relaxation time is observed to show a stronger than linear dependence on the sample age, suggestive of hyper-aging dynamics. We show that these features are consistent with non-monotonic steady-state shear stress/shear rate flow curves characterized by a local stress minimum. When a steady shear flow is suddenly ceased, and the total imposed sample strain is held constant, both materials show an initial relaxation of the shear stress followed by a period of shear stress buildup, resulting in a local minimum in the evolution of shear stress with time. For the clay dispersion, the intensity of these effects increases with higher pre-shear rates, whereas for the silica gel, the effects are largely independent of the pre-shear rate. We also propose a simple time-dependent linear Maxwell model, which qualitatively predicts the experimentally observed trends in which the shear stress build-up is directly related to a monotonic increase in the elastic modulus, giving keen insight into this peculiar phenomenon.

# I. Introduction

The thermodynamic drive to attain an equilibrium state is an inherent tendency of any material [1]. However, in a specific class of soft materials, the constituents can become arrested in physical (owing to high density) or energetic (due to strong attractive interactions) 'cages' that limit the material elements' access to their entire phase space [2-4]. Such materials cannot reach their equilibrium state at room temperatures over practical time scales and are structurally arrested; they are thus commonly referred to as *soft glassy materials*. While unable to reach a global equilibrium state, the internal microstructure of soft glassy materials with thermally driven constituents undergoes continuous evolution and passes through a series of progressively lower local energy states as a function of time, a phenomenon known as physical aging [4-7]. The external application of a deformation field breaks down the microstructure that has developed during aging, a phenomenon known as *rejuvenation* or *letherization* [8-10]. Many soft materials, such as concentrated suspensions and emulsions [11, 12], colloidal gels [2, 7, 13, 14], industrial slurries [15], pastes [16, 17], foams [11], microgels [6, 18-20], waxy crude oil [21, 22], *etc.*, belong to this category of complex fluid. For a specific deformation history, the observed rheological behavior of soft glassy materials is controlled by the interplay between aging and rejuvenation. This interaction results in a number of unusual and apparently contradictory phenomena, including viscosity bifurcation as the level of imposed shear stress is varied [11, 15, 16, 23-26], time-dependent yield stress [25, 27-31], transient and/or steady state shear banding [12, 27, 32-37], delayed yielding [14, 38-41], delayed solidification [42, 43], over-aging [19, 44-46], as well as the presence of significant internal residual stresses [20, 46-54] and/or strains [6, 55], amongst others. In this work, we investigate the intriguing phenomenon of shear stress build-up under constant strain conditions in two model soft glassy materials.

In a number of soft glassy materials, the shear stress introduced by preceding deformation protocols does not relax completely after flow cessation, leading to residual stresses in the system. The specific nature of the relaxation behavior and the magnitude of the residual stress depend on the characteristics (composition, microstructure, etc.) of the material and the details of the entire prior shearing history. Ballauff et al. [47] studied colloidal hard-sphere glasses and observed that a higher concentration of the particles and lower pre-shear rates induced greater residual shear stress. They also performed molecular dynamics simulation and analyzed the resulting rheological behavior theoretically through mode-coupling theory using an integration-through-transients (MCT-ITT) framework, corroborating their experimental findings on hard sphere glasses. They proposed that long-lived memory effects influence the magnitude and evolution of residual stresses within the system. Interestingly, Fritschi et al. [48] also made similar observations for colloidal hard-sphere glasses by employing the MCT-ITT framework in combination with Brownian dynamic simulations. Mohan et al. [51, 52] and Lidon et al. [20] investigated microgel pastes with very different microstructures compared to hard-sphere glasses. In stress relaxation after steady shear, they observed a similar monotonic decrease in residual stress with an increasing pre-shear rate. Mohan et al. [51, 52] observed the initial stress relaxation in their paste-like systems to be very rapid, which they attributed to local ballistic particle motion. Subsequently, the stress relaxation becomes slower, which they ascribe to weak rearrangement of local angular distortions. At longer times, they reported a finite residual stress owing to the arrested structure of the constituents. In a recent molecular dynamics and mesoscopic simulation study, Vasisth et al. [54] observed a similar monotonic decrease in residual stress with increased pre-shear rate in non-Brownian suspensions of soft repulsive spheres. Through analysis of spatial mesoscale maps of active (yielded) sites and local stress distributions, they reported that the majority of stress relaxation occurs in the system due to the formation (during the period of pre-shear) of new rate-dependent active sites in local regions adjacent to the existing yielded sites.

The observation of residual stresses noted above suggests extremely slow thermally-driven relaxation of shear stress in the material over time while maintaining a monotonic rate of decrease that results in an eventual plateau in the residual stress. However, recently, it has been reported that some soft glassy materials display unusual *non-monotonic* shear stress relaxation behavior. The first experimental report of such behavior was by Negi and Osuji [53], who studied aqueous Laponite® RD gel and found that the rates of shear stress relaxation and subsequent stress build-up increase with an increase in the pre-shear rate, while the magnitude of the eventual residual stress decreases. However, their work does not identify the observation explicitly as a shear stress build-up phenomenon. More recently, Hendricks et al. [49] reported shear stress build-up behavior for three different gel-forming supramolecular assemblies. Following a pre-shearing deformation, stress relaxation measurements showed that there was an initial decrease in the sample shear stress, which then increased at intermediate times before passing through a maximum and eventually decaying away completely. Hendricks et al. attributed the observed shear stress build-up over intermediate time scales to bond formation between partially aligned elastic domains. For Boehmite gels prepared in aqueous nitric acid, Sudreau et al. [46] reported two different stress relaxation behaviors depending on a critical pre-shear rate. Above the critical shear rate, they observed shear stress build-up behavior similar to that of Hendricks et al. [49]; again, this evolution was non-monotonic, and after a period of stress build-up, the shear stress eventually relaxed completely. On the other hand, for pre-shear rates lower than the critical shear rate, they observed shear stress build-up combined with the presence of permanent residual stress in the system. The magnitude of this residual shear stress increased with decreasing pre-shear rate.

Joshi [50] was the first to propose the possibility of shear stress build-up theoretically through a structural kinetic model framework that considered an increase in the material shear modulus with time. Interestingly, a spontaneous increase in the shear stress when the total imposed sample strain is held constant suggests the possibility of violating the second law of thermodynamics [56]. To address this issue, Joshi, in subsequent work [57], showed that when the rejuvenation term in the structural kinetic model is only a function of the viscous (irreversible) contribution to the total rate of shear deformation, the second law of thermodynamics does not get violated even though the resulting constitutive model predicts a non-monotonic stress relaxation. Joshi [57] also presented a simple structural kinetic model, which rationalized the experimental data of Hendricks et al. [49] qualitatively. Interestingly, the Soft Glassy Rheology (SGR) model [58], when studied below the glass transition temperature (i.e. at effective noise temperatures $x < 1$) shows an increase in $G'(\omega, t)$ with time for a brief period after shear rejuvenation as it approaches an asymptotic constant value of the elastic modulus. However, in stress relaxation subsequent to a step strain deformation, the SGR model does not show any stress build-up [58]. Recently, Ward and Fielding [59] modified the SGR model to allow for steady-state shear banding in their simulation. They demonstrated how shear stress build-up might also arise when localized shear bands are present in the steady rate-dependent material microstructure that is present in the system before the flow stops.

While stress build-up behavior has now been observed in a few complex fluid systems, several issues require careful examination, especially concerning soft glassy materials that undergo physical aging. The first question involves the effect of the characteristic features of physical aging, such as the rate of increase in the age-dependent linear viscoelastic modulus and the change in the microstructural relaxation time with sample age. Another question is how the degree of pre-shear affects the subsequent evolution of these materials. Additionally, understanding the typical features of the steady-state flow curve for such materials is also important, as well as determining how long the effects of shear stress build-up persist in the material after the cessation of the pre-shear step. In addition, prior to the observation of stress buildup under constant strain conditions, that is, before the steady state pre-shear stops, it is important to deduce whether the flow field in the shear cell is homogeneous or shear-banded. To explore these questions, we study two distinct colloidal gel systems, namely, aqueous dispersions of Laponite® RD (a discotic clay system) and Ludox® (a silica-based system). Interestingly, both the systems exhibit similar aging dynamics under quiescent conditions: both the relaxation time and the elastic modulus increase with the age of the sample (i.e., with the elapsed time since the cessation of a previous rejuvenation/letherization step). Both of the materials also show shear stress build-up behavior following cessation of a strong steady pre-shearing step, which we analyze and quantify through the evolution in their macroscopic rheological characteristics. These two measurements then motivate a simple phenomenological model that can capture the key experimental observations.

## II. Materials and Methods:

## II.1 Sample preparation:

In this work, we used two distinct model systems: a 30 day old aqueous dispersion of 3.5 wt.% Laponite® RD (BYK Ltd.) and 2 day old aqueous dispersion of Ludox® TM-40 (Sigma-Aldrich) colloidal silica. Because Laponite samples in water undergo a long, slow irreversible *chemical* aging associated with space spanning network formation of the discotic clay sheets [60, 61], control of the entire sample history is important. The as-received sample (dry white powder) of Laponite® RD clay was first dried at 120 ℃ for 4 hours to remove any residual moisture. Subsequently, the predetermined amount of clay was gradually added to ultrapure water and mixed for 30 minutes using an ULTRA TURRAX® TM-25 homogenizer operating at 9500 rpm. The resulting Laponite® RD dispersion was hermetically sealed in a polypropylene bottle and preserved in an undisturbed state for 30 days before experimentation. Rheological experiments performed over a period of 2-3 days after this initial irreversible chemical conditioning step can be used to probe the effects of physically aging (with characteristic time scales of minutes to hours) with respect to a well-conditioned (but very slowly varying) base reference state.

The Ludox® gel was prepared by diluting a stock solution of Ludox® TM-40 colloidal silica in ultrapure water containing 10 wt.% NaCl as per the methodology prescribed by Kurokawa et al. [32]. In this work, we use a dispersion having a weight ratio of 6:13 between the stock silica and saline water. After homogenization for 2 minutes at 9500 rpm using the ULTRA TURRAX® TM-25, the Ludox® gel dispersion was stored in a tightly sealed polypropylene bottle for 2 days before carrying out the experiments. To prepare both dispersions, ultrapure Millipore water with a resistivity of 18.2 MΩ.cm was used. Hereafter, for compactness, we refer to the 30 day old Laponite® RD sample as the clay dispersion and the 2 day old sample Ludox® as the silica gel.

## II.2 Rheological Experiments:

The rheological experiments described in this work have been carried out using an Anton Paar MCR-501 stress-controlled rheometer. We used a concentric cylinder geometry (cup diameter: 28.915 mm, bob diameter: 26.650 mm, and bob length: 40.020 mm) having serrated walls to minimize/eliminate any effects of wall slip. After loading the sample in the rheometer geometry, both the materials were subjected to shear melting or letherization at high shear rates ($100\ s^{-1}$ for the Laponite® RD dispersion and $1000\ s^{-1}$ for the Ludox® gel) for 1000 s to eliminate any effects of prior deformation/loading history. For notational clarity, we re-zero the time origin at the instant of cessation of shear melting and plot the subsequent experimental data. In this work, we carry out the following experiments subsequent to shear melting:

1. Physical aging: The samples were subjected to time sweep experiments at different angular frequencies ($\omega = [0.5 - 10]$ rad/s) in the linear viscoelastic domain to investigate the temporal evolution of viscoelastic properties over a period of one hour. To understand the effect of the pre-shear rate on aging, the systems were also rejuvenated at different shear rates in the range of 100 – 1000 $s^{-1}$ prior to the time sweep experiments. During the measurements of physical aging, the oscillatory strain amplitudes were set at 1% for the Laponite® RD dispersion and 0.1% for Ludox® gel. In a small amplitude oscillatory experiment, dynamic moduli are computed from the measured deformation over a period of (at least) one cycle, corresponding to an experimental acquisition time $t_{\exp} = 2\pi/\omega$, where $\omega$ is the angular frequency. For time-dependent materials, it is necessary to ascertain that the change in the material properties is not significant over a single period of oscillation. This change can be quantified by evaluating a relevant time scale for the rate of evolution in the material property being measured. For the case of a time-evolving storage modulus being measured at a single, constant imposed frequency, we can define an appropriate mutation time scale [62] at any instant of time $t$ as: $\tau_{\mathrm{mu}} = G'(\omega)/(\partial G'/\partial t)_\omega = (\partial \ln G'(\omega)/\partial t)^{-1}$. Since the evolution in the elastic modulus of a physically aging sample is a direct measure of the build-up in the internal microstructure and is commonly plotted on log-log axes, we define the *logarithmic rate of microstructural evolution* (LROME), as $(\partial \ln G'(\omega)/\partial \ln t)$; or by expanding the $\partial \ln t$ term we find LROME $\equiv t/\tau_{\mathrm{mu}}$. This establishes a direct connection between $\tau_{\mathrm{mu}}$ and a measure of microstructural evolution.

2. Mours and Winter [62] argued that the evaluation of time- and frequency-dependent dynamic moduli using conventional rheometric techniques such as SAOS is appropriate when the magnitude of the *sample mutation numbers* defined as $N'_{\mathrm{Mu}} \equiv \frac{t_{\exp}}{\tau_{\mathrm{mu}}} = \frac{2\pi}{\omega G'}\left|\frac{\partial G'}{\partial t}\right| \ll 1$. A similar criterion can of course also be defined in terms of the loss modulus (although in physically aging materials at higher ages this typically decreases with sample age) as $N''_{\mathrm{Mu}} = \frac{2\pi}{\omega G''}\left|\frac{\partial G''}{\partial t}\right|$. Mours and Winter recommend that experiments are performed with sufficiently high oscillatory test frequencies so that the sample mutation numbers are kept below $N_{\mathrm{Mu}} \leq 0.1$. More sophisticated techniques using optimally-windowed chirps and Gabor transforms can also be used if it is desired to monitor the evolution of the entire linear viscoelastic spectrum of rapidly aging systems [15, 63], however this is not necessary for the present samples. We monitor the mutation numbers at different sample age times and oscillatory test frequencies and the majority of results presented here correspond to $N'_{\mathrm{Mu}} \leq 0.1$. However, for some sufficiently low frequencies and early times (less than 100 $s$ following cessation of shear after shear melting) the values of mutation numbers are observed to be higher but remain below unity for all conditions (refer to the supplementary information Figs. S1 and S2).

3. Steady-state behavior: Steady-state flow curves were generated by applying a constant shear rate (CSR) to the samples for 1000 s in a stepwise manner, decreasing the shear rate monotonically from 1000 s$^{-1}$ to 100 s$^{-1}$. Similarly, we also carry out equivalent experiments by applying a constant shear stress (CSS) to the samples for 3600 s and decreasing the applied stress stepwise from 60 Pa to 20 Pa for Laponite RD® and from 8 Pa to 1 Pa for Ludox®.
4. Rheological behavior at different waiting times: The influence of waiting time ($t_w$) since the cessation of the shear melting (rejuvenation step) on material properties was monitored by subjecting the samples to a number of different viscometric test protocols. After different waiting times, we performed strain sweep experiments at a relatively high test frequency of 10 rad/s (corresponding to $t_{exp} < 1$ s), which led to an accurate estimation of the static yield stress of the sample. Creep/recoil experiments at stresses of 1 and 0 Pa on Laponite RD® and at 0.5 and 0 Pa on Ludox® were performed to determine the dependence of the material relaxation time on sample age or waiting time. Over the waiting period ($t_w$), we subjected the samples to an oscillatory field with a strain amplitude of 0.1% and $\omega$ =0.63 rad/s angular frequency.
5. Cessation of steady shear: The samples were subjected to steady shear flow over a range of different shear rates (also called pre-shearing rates, $\dot{\gamma}_{PS}$), and the approach to a steady state was ensured. At a certain point (corresponding to a pre-shearing time of $t_{PS}$ =1000 s), the flow was suddenly stopped ($\dot{\gamma} = 0$). Subsequently, the relaxation/evolution of shear stress was measured for one hour.

To prevent the evaporation of water during experiments, the free surfaces of the samples were coated with a thin film of low-viscosity silicone oil. All the experiments were conducted at a constant temperature of 25 ℃.

## III. Experimental Results:

### III.1 Viscoelasticity and Physical Aging

Before commencing experiments, both of the test materials were shear melted or letherized with the goal of erasing residual effects of prior deformation histories. Subsequently, to study the physical aging behavior, the materials were subjected to an oscillatory flow field at different angular frequencies for one hour, and the time-dependent evolution of both the elastic ($G'(t;\omega)$) and viscous ($G''(t;\omega)$) moduli were monitored. The corresponding behaviors for the clay dispersion and silica gel are shown in Figs. 1(a) and 1(b) for a pre-shear rate $\dot{\gamma}_{PS}$= 1000 s$^{-1}$ ($t_{PS}$ = 1000 s). In the clay dispersion, the viscoelastic moduli show strong initial dependence on the angular frequency, and the initial rate of increase of $G'$ with time is much faster for the lowest probing angular frequency (0.5 rad/s). Artifacts associated with finite sample mutation numbers may play a role here (refer to the supplementary information Figs. S1 and S2).

However, after this initial frequency dependence, at larger times ($t \geq 300$ s), all the evolution curves of $G'(t)$ for the clay dispersion merge and evolve at the same rate. The logarithmic rate of microstructural evolution (LROME) is thus frequency independent and decreases monotonically with time and eventually attains a constant value over the explored period of the experiment.

Interestingly, for the silica gel system, the evolution behavior of $G'$ remains independent of $\omega$ from the very start. After the initial fast increase in the age-dependent elastic modulus, eventually, the age-dependent elastic modulus $G'$ for each of the systems shows a power law dependence on $t$. In contrast, after an initial increase with time, the age-dependent loss modulus $G''(t; \omega)$ of both systems starts decreasing, with the loss modulus of the clay dispersion showing a faster decrease in comparison to the silica gel. Interestingly, the age-dependent loss modulus $G''$ of the clay dispersion shows weaker dependence on $\omega$ than the silica gel. This generic kind of behavior has been reported for many aging soft glassy materials [2, 15, 25].

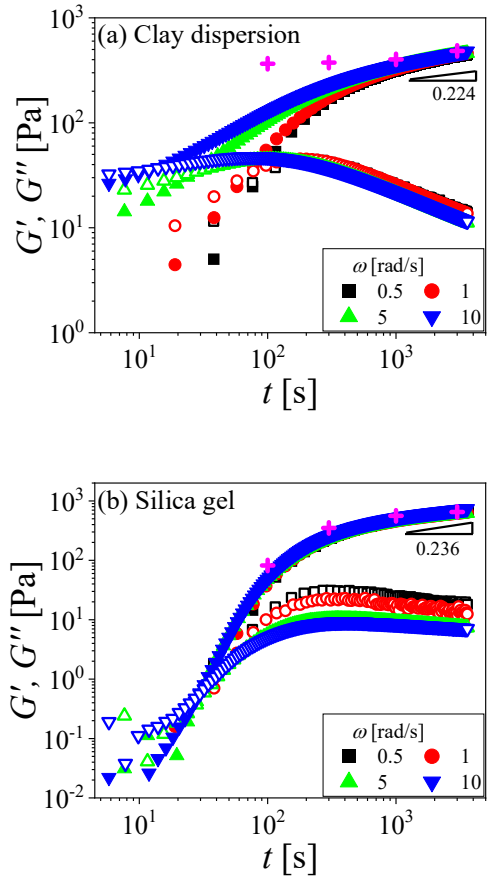

Figure 1: Evolution of elastic ($G'$) (filled symbol) and viscous moduli ($G''$) (open symbol) with time ($t$) at different angular frequencies ($\omega$) (values are given in the legend) for (a) clay dispersion at 1% strain amplitude and (b) silica gel at 0.1% strain amplitude after pre-shearing the systems at 1000 $s^{-1}$. The plus (+) symbols (magenta colour) are the values of $G'$ used for vertical shifting the creep compliance data for the time – waiting-time superposition (TWTS), shown in Fig 8 (a, b).

In Fig. 2, we plot the dynamic moduli as a function of angular frequency at three different waiting times ($t_w$) for clay dispersion (Fig. 2(a)) and silica gel (Fig. 2(b)). At short waiting times ($t_w$ = 100 s), the elastic modulus ($G'(\omega; t_w)$) of the aging clay dispersion shows a strong dependence on the angular frequency ($\omega$) and increases with increasing $\omega$, whereas for the silica gel, $G'(\omega; t_w)$ remains essentially independent of $\omega$. As the waiting time increases, the dependence of $G'(\omega; t_w)$ of the clay dispersion on $\omega$ also diminishes, and at larger waiting time, it becomes independent of $\omega$ as also predicted by the SGR model [58]. On the other hand, $G''(\omega; t_w)$ of silica gel decreases with $\omega$ while that of clay dispersion remains independent of $\omega$.

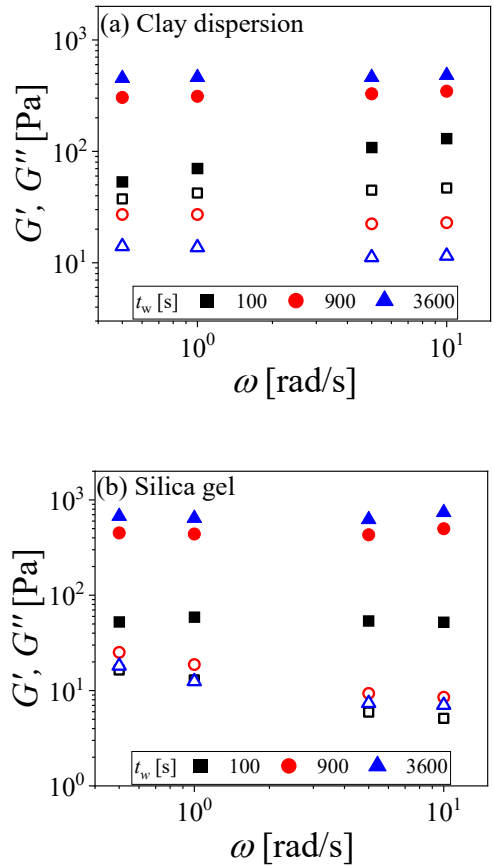

Figure 2: Variation of the age-dependent elastic moduli ($G'(\omega; t_w)$; filled symbol) and viscous moduli ($G''(\omega; t_w)$; open symbol) with angular frequency ($\omega$) at different waiting times ($t_w$) (values are shown in the legend) for (a) the clay dispersion at 1% strain amplitude and (b) silica gel at 0.1% strain amplitude.

## III.2 Time-dependent yield stress, Viscosity bifurcation, and flow curve

Figs. 1 and 2 clearly show an increase in $G'$ for both the materials as a function of time. We applied an oscillatory flow field with increasing strain amplitude on these systems at a high frequency $\omega = 10$ rad/s (corresponding to $t_{\text{exp}} \leq 1$ s) to determine the static yield stress ($\sigma_y$) at various waiting times ($t_w$) since the cessation of shear melting. The corresponding results are shown in Fig. 3(a) and (b) for the clay dispersion and silica gel.

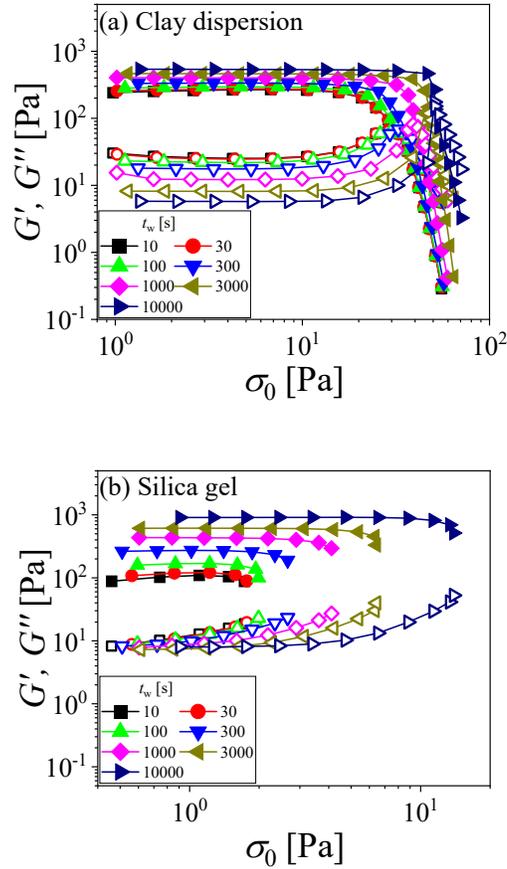

Figure 3: Variation of age-dependent $G'$ (closed symbol) and $G''$ (open symbol) with increasing magnitude of oscillatory stress at $\omega$ = 10 rad/s for (a) clay dispersion and (b) silica gel. Symbols in the legend represent different waiting times ($t_w$) at which experiments were performed.

Within the linear viscoelastic domain, both the storage and loss moduli remain constant with respect to the applied stresses (although, of course, they show a slow time-dependent aging characterized by the power-law logarithmic rate of microstructural evolution). As the yield stress is approached, $G'$ starts to show a gradual decrease while $G''$ undergoes a noticeable increase in both materials. At the point of yielding, $G'$ shows a more sudden decrease and eventually decreases below the value of $G''$. Although there are many possible metrics for defining the onset of nonlinearity and yielding in a thixotropic yield stress material [64, 65], we identify the critical value of the applied stress at which the elastic (in-phase) and dissipative (out-of-phase) contributions to the cyclic deformation are equal, as a measure of the static yield stress ($\sigma_y(t_w)$) of the aging material. In the clay dispersion, dynamic moduli, as recorded by the rheometer from the first harmonic of the stress response, are measurable even beyond the yield point and are also shown in Fig. 3(a). For the silica gel, the dynamic moduli and stress decrease sharply and become immeasurably small after the yield point (the elastoplastic sample rapidly becomes a viscoplastic liquid beyond the critical stress); hence, it has been omitted from Fig. 3(b). The corresponding values of the age-dependent static yield stress $\sigma_y(t_w)$ are shown as a function of waiting time ($t_w$) in Fig. 4 for both materials. Interestingly, $\sigma_y$ for both materials remains approximately constant for small wait times and then shows a continuous increase as a function of waiting time. For both systems, the time up to which $\sigma_y$ remains constant is around 100 s. In this regime, the static yield stress values for clay dispersion and silica gel are 33.2 Pa and 1.8 Pa, respectively.

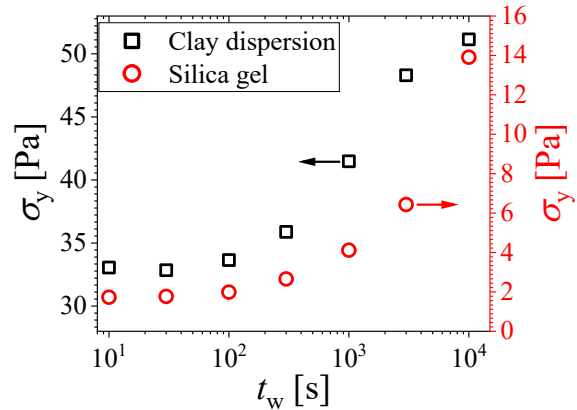

Figure 4: Evolution of the static yield stress ($\sigma_y$) as a function of waiting time ($t_w$) for the clay dispersion (open black squares), and the silica gel (open red circles). The static yield stress is measured at the point of intersection of $G'$ and $G''$ from the data presented Fig. 3. The arrows are a guide for the eye.

Next, we conducted a series of creep experiments on the clay dispersion and silica gel over a range of steady imposed stresses, to identify the critical flow stress associated with viscosity bifurcation. In addition, this procedure aids in the determination of the nature of the steady-state shear stress–shear rate flow curve, particularly the minimum shear rate consistent with sustained homogeneous shearing flow. These experiments involve first aging a rejuvenated sample for a predetermined time ($t_w = 100$ s for clay dispersion and $t_w = 5$ s for silica gel in the present study) in order to allow a consistent state of internal microstructure to be established and then applying a constant shear stress over a range of values spanning the static yield stress determined above.

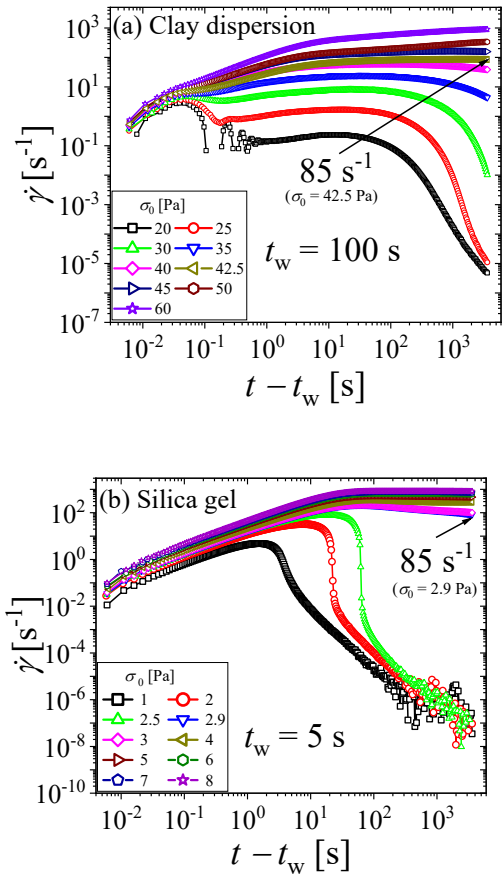

Figure 5: Evolution of shear rate with time at different applied shear stresses (values are shown in the legend) for (a) clay dispersion and (b) silica gel. During the waiting time, an oscillatory shearing field with 0.1% strain amplitude and 0.63 rad/s angular frequency was applied.

In Fig. 5, we plot the corresponding evolution of the instantaneous shear rate as a function of time under the application of constant stress. We also plot the viscosity bifurcation curves for both systems in terms of the transient shear viscosity ($\eta^+ = \sigma_0/\dot{\gamma}(t-t_w)$) for the different applied stresses in the supplementary information Fig. S3. It can be seen that above a critical threshold that is specific to each material ($\sigma_c = 42.5$ Pa for the clay dispersion and $\sigma_c = 2.9$ Pa for silica gel), a steady flowing state with a constant steady shear rate (and viscosity) is achieved. Below this threshold stress, however, the shear rate in the sample increases initially before passing through a maximum and gradually decaying toward zero (and the corresponding transient viscosity thus diverges), leading to a cessation of the deformation; i.e. the sample creeps but does not ever approach a state corresponding to steady viscous flow. The corresponding threshold stress has been termed a critical threshold stress associated with viscosity bifurcation [23]. The observed response is attributed to the interplay between aging (i.e. the logarithmic rate of microstructure formation) and the rate of sample rejuvenation (breakdown) under applied stress [50, 66]. Rejuvenation prevails when the applied stress exceeds the threshold stress, leading to a steady state. On the other hand, for stresses below the threshold, the dominance of physical aging leads to a build-up of sample microstructure, a decrease in the rate of creep, and eventually flow cessation [38, 67]. We draw attention to the differences in yield stress values obtained from oscillatory shear experiments (Fig. 4) and viscosity bifurcation experiments (Fig. 5). In the limit of small waiting times, the minimum static yield stress recorded is 33.2 Pa for clay dispersion and 1.8 Pa for colloidal silica dispersion. In both systems, the yield stress begins to increase for waiting times exceeding 100 s. Viscosity bifurcation experiments were conducted at a waiting time of 100 s for clay dispersions and 5 s for colloidal silica dispersions, resulting in corresponding yield stress values of 42.5 Pa and 2.9 Pa, respectively. Given that these two experimental protocols differ, with the creep experiment representing a limiting case of a zero-frequency test, while oscillatory experiments are conducted at nonzero frequencies, subjecting the material to a specific strain amplitude for a finite duration. Additionally, the creep and oscillatory experiments were performed on different but closely spaced dates, which may have introduced some irreversible changes despite shear melting. Notwithstanding these variations, the primary yield stress values obtained from both protocols for both systems are comparable. Based on this, we report the minimum static yield stress as $38 \pm 5$ Pa for the clay dispersion and $2.4 \pm 0.5$ Pa for the colloidal silica dispersion.

The observed time-dependent increase in yield stress and the distinct occurrence of viscosity bifurcation may have significant consequences. As discussed in the introduction section, soft glassy materials may show a non-monotonic flow curve such that steady-state shear stress - shear rate dependence has a negative slope in the low shear rate region [25, 50]. However, it is impossible to deduce the non-monotonic nature of the flow curve through conventional rheometric protocols. Typically, it has been observed that when a constant external shear rate is applied in the region where the dependence of shear stress on the imposed shear rate has a negative slope, the sample undergoes steady-state shear banding [12, 24, 33, 35, 68-70]. However, without velocimetric instrumentation coupled to the rheometer, such as a particle image velocimetry, confocal microscopy, etc., one cannot explicitly establish if a material sample has shear banded or not. However, there are implicit ways to ascertain whether the material is characterized by a non-monotonic flow curve. As suggested in the literature [25, 50], observation of viscosity bifurcation and the evolution in the time-dependent elastic modulus, as well as the yield stress, is consistent with the non-monotonic nature of the flow curve. The stable region of the flow curve is where shear stress increases with shear rate. The viscosity bifurcation experiment helps clearly identify this stable region. In a creep experiment, where a constant stress is applied, the system eventually reaches a steady-state shear rate. The lowest steady-state shear rate observed in such an experiment represents a stable point on the flow curve, as shown in Fig. 5. Below this point, the flow curve may become non-monotonic, meaning that an increase in shear stress does not necessarily lead to an increase in shear rate, potentially leading to unstable or discontinuous flow behavior. Although the critical stresses are markedly different for each sample they correspond to shear rates close to 80 – 100 $s^{-1}$ for both the clay dispersion and silica gel.

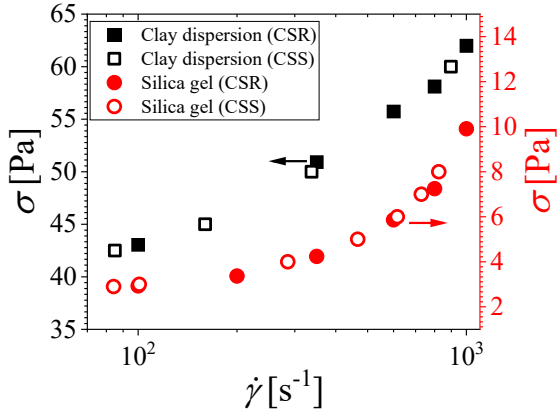

Figure 6: Steady-state flow curve of clay dispersion (squares), and silica gel (circles). The closed and open symbols represent the imposed constant shear rate (CSR) and constant shear stress (CSS) deformation fields respectively.

In Fig. 6, we plot the corresponding steady-state shear stress – shear rate flow curves for both the clay dispersion and the silica gel. In these plots, data from both constant shear stress (CSS) creep protocols (i.e. viscosity bifurcation experiments) and constant shear rate (CSR) experiments have been plotted. Remarkably, the data from both experiments show an excellent overlap, highlighting the consistency and homogeneity of the internal deformation field over this imposed range of stresses and shear rates. The minimum shear rates associated with the steady shear flow that we obtain in the viscosity bifurcation experiments mark the boundary of homogeneous flow; any externally-imposed value of shear rate below this threshold can cause the emergence of steady state shear banded states [33, 35-37] and we thus avoid this regime in our subsequent experiments.

### III.3 Age-Dependent Viscoelastic Relaxation Time

Next, we investigate the time-dependent evolution of the viscoelastic relaxation time in both of these aging elastoviscioplastic materials. Accordingly, subsequent to shear melting, constant shear stress is applied to both materials at different waiting times ($t_w$), and the resulting evolution in the creep compliance is measured. Since it is known that the relaxation time ceases to show an age-time dependence for creep stresses above the yield stress [6, 25, 38, 67, 71], we performed the creep measurements with the stress of 1 Pa for clay dispersion and 0.5 Pa for silica gel. While the shear melting erases memory associated with prior deformation histories, the creep response of both aging samples shown in Fig. 7 is affected by the waiting time allowed for restructuration. As expected, the shorter the waiting time, the more pronounced the effect of shear melting on the sample, which gradually diminishes as the waiting time increases. To address this, we apply zero stress to both materials at identical waiting times and measure the strain recovery. Subsequently, we obtain the actual strain by adding strain recovered under no-stress conditions to the strain induced during creep as discussed elsewhere [18].

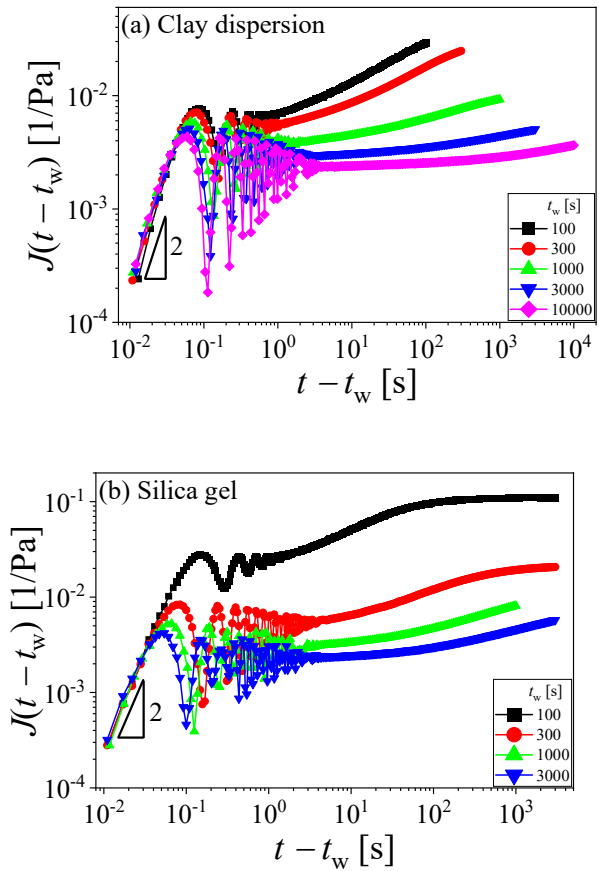

Figure 7: Evolution of creep compliance $J(t - t_w)$ with creep time $(t - t_w)$ at different waiting times for (a) clay dispersion (at a stress $\sigma_0 = 1$ Pa) and (b) silica gel (at a stress $\sigma_0 = 0.5$ Pa). Legends show the values of $t_w$ at which creep experiments were performed.

We plot the corrected compliance for both materials in Fig. 7 (a, b). It can be seen that when compared at any fixed value of the creep time $(t - t_w)$, an increase in the sample age $(t_w)$ causes a decrease in creep compliance. Moreover, the rate of evolution of creep compliance also gets weaker with an increase in waiting time. Such compliance behavior has been commonly observed in physically aging materials, including soft glassy materials [15, 18, 72, 73]. Interestingly, in Fig. 7(a, b), it can also be observed that in both experimental systems, the compliance data exhibits creep ringing at shorter timescales of (1) s, and a quadratic dependence on creep time at very short timescales $O(10^{-2})$ s. In a stress-controlled rheometer, this behavior is observed due to the coupling between the sample elasticity and instrument inertia [16, 71, 74, 75]. The elapsed time to the first overshoot in the evolving compliance $J(t - t_w, t_w)$ and the period of the damped inertia-elastic oscillations both decrease as the samples age and stiffen.

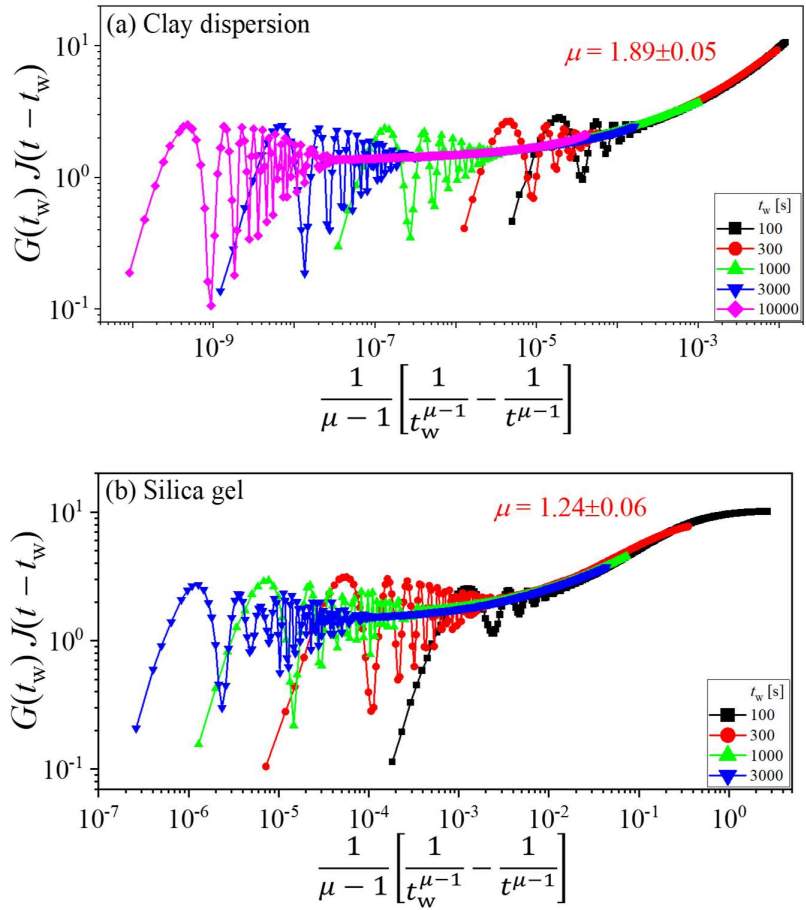

Figure 8: Time – waiting-time superposition (TWTS) plots for (a) clay dispersion and (b) silica gel, respectively. Legends show the value of waiting times $(t_w)$ at which creep experiments were performed.

As shown in Fig. 7(a, b), the creep compliance for both materials not only shows dependence on time elapsed since the application of step stress (or creep time, $t - t_w$) but also on waiting time ($t_w$), leading to $J = J(t - t_w, t_w)$. This violates time–translational invariance (TTI), a necessary condition for applying linear viscoelastic principles to both materials, including the Boltzmann superposition principle [15, 76-79]. This issue can be addressed by transforming the material clock from the real-time domain ($t$) to the effective time domain ($\xi(t)$) defined as [72],

$$\xi(t) = \tau_{ETD} \int_0^t \frac{dt'}{\tau(t')}. \qquad (1)$$

According to Eq. (1), the material clock is adjusted in such a fashion that the relaxation time of the mutating material, which depends on the elapsed wall-clock time in the real-time domain ($i.e. \tau = \tau(t')$), takes a constant value of $\tau_{ETD}$ in the effective time domain. Consequently, in the transformed domain, the viscoelastic material obeys effective time translational invariance [67]. The Boltzmann superposition principle in this effective time domain can then be expressed as:

$$\gamma(\xi) = \int_{-\infty}^{\xi} J(\xi - \xi_w) \frac{d\sigma}{d\xi_w} d\xi_w, \qquad (2)$$

where $\xi_w = \xi(t_w)$ and $J(\xi - \xi_w)$ is the creep compliance, and $\xi - \xi_w$ is the effective time elapsed since the application of the deformation field. However, to express the effective time-domain given by Eq. (1), it is necessary to have prior information about the dependence of the material relaxation time on sample age (i.e. on the elapsed wall-clock time). Usually, a certain functional form of relaxation time is assumed, and using Eqs. (1) and (2), it is confirmed that the creep compliance is indeed just a function of the transformed time variable $\xi - \xi_w$. In other words it is determined whether the creep curves at different ages show superposition in the effective time domain. In general, in the case of soft glassy materials, the relaxation time has been reported to display a power law dependence [15, 18, 25, 73] on wall-clock time (sample age) given by:

$$\tau = A_1 \tau_m \left(\frac{t}{\tau_m}\right)^\mu, \qquad (3)$$

where $A_1$ is a constant and $\tau_m$ is the characteristic timescale associated with the microstructural reorganization of the soft glassy material constituents. In this expression, the power law coefficient $\mu$ is given by: $\mu = \frac{d\ln\tau}{d\ln t}$. The material is classified as hyper-aging if $\mu > 1$, simple aging if $\mu = 1$, and sub-aging if $\mu < 1$ [67, 80]. Incorporating this general expression for the characteristic relaxation time into Eq. 1 and considering $\tau_{ETD} = \tau_m$, we get:

$$\xi - \xi_w = \tau_m \int_{t_w}^t \frac{dt'}{\tau(t')} = \frac{\tau_m^\mu}{A_1} \frac{1}{\mu - 1} \left[\frac{1}{t_w^{\mu-1}} - \frac{1}{t^{\mu-1}}\right] \qquad (4)$$

Eq. (4) suggests that, for materials showing hyper-aging dynamics ($\mu > 1$), $\xi - \xi_w$ approaches an asymptotic constant value in a limit of $t - t_w > \mu t_w/(\mu - 1)$ [15, 76]. In the effective time domain, since the creep compliance is only function of $\xi - \xi_w$, it is thus to be expected that the compliance will approach a plateau value in the limit of creep times much longer than the waiting time. Because of the strongly nonlinear mapping between the effective time and the laboratory time domain, even in the real (laboratory) time-domain the creep compliance is expected to slowly approach a plateau for materials undergoing hyper-aging dynamics [15, 76]. Furthermore, as the waiting time ($t_w$) increases, the elapsed creep time at which the compliance plateaus out is observed to increase. Both of these aspects are clearly observed in Fig. 7(b); for the smallest waiting times of 100 and 300 s, we can indeed observe that the compliance becomes constant for $t - t_w > \mu t_w/(\mu - 1)$ and also that the creep time at which the compliance curve plateaus out also increases with the sample age. In Fig. 8 (a, b), we plot the creep compliance normalized by the elastic modulus of the aged sample, i.e. the product $G(t_w)J(t - t_w)$ versus the elapsed effective time $\frac{A_1}{\tau_m^\mu}(\xi - \xi_w) \equiv \frac{1}{\mu-1}\left[\frac{1}{t_w^{\mu-1}} - \frac{1}{t^{\mu-1}}\right]$. It can be seen that (after all of the nonlinear inertioelastic oscillations decay away) the compliance curves superimpose on each other, consistent with master curves characterized by $\mu = 1.89 \pm 0.05$ for the clay dispersion and by $\mu = 1.24 \pm 0.06$ for the silica gel. As expected, the scaled creep compliance shows a smooth asymptotic approach to a distinct material-specific plateau value for higher values of $\xi - \xi_w$. This general behavior is a powerful demonstration that both materials show hyper-aging dynamics ($\mu > 1$) and validates the application of the Boltzmann superposition principle in the effective time domain.

### III.4 Stress Build-up

The discussion until now clearly underlines the following aspects: the elastic modulus and yield stress for both materials increase as a function of time (i.e., sample age), and the characteristic relaxation time of both materials also increases faster than linearly with respect to time. For such a scenario, it has been proposed that the steady-state shear stress–shear rate flow curve shows a non-monotonic flow curve [50]. Fig. 4(a) also clearly shows features consistent with such yield stress behavior, while the viscosity bifurcation experiments help identify the minimum imposed shear rate above which the flow field remains homogeneous ($\approx 85$ s$^{-1}$ for both systems). Next, we study how the viscoelastic stress relaxation in such materials following the cessation of steady shear flow is affected by the magnitude of the shear rate. To ensure conditions consistent with homogeneous flow during the period of steady shearing, we employ a minimum pre-shear rate for both materials of $\dot{\gamma}_{PS}$ = 100 s$^{-1}$ with the maximum value set at $\dot{\gamma}_{PS} = 1000$ s$^{-1}$.

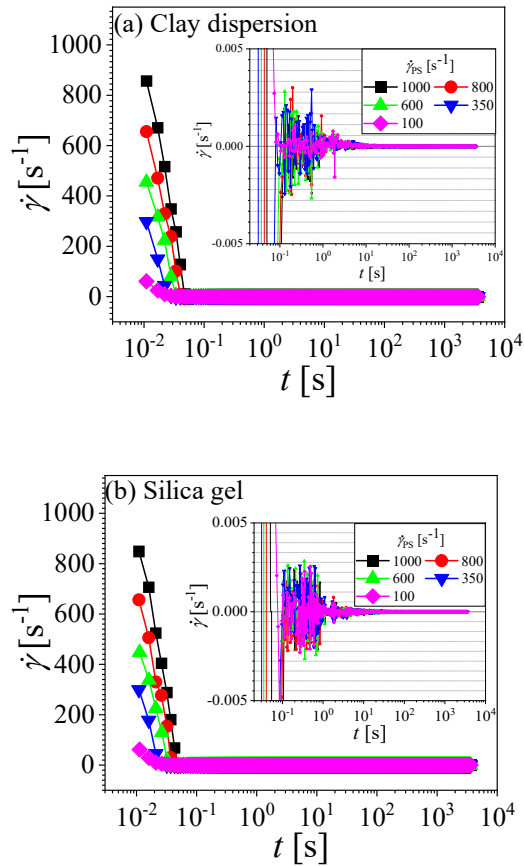

Figure 9: Variation of the instantaneous shear rate during the cessation of flow for (a) clay dispersion and (b) silica gel. The legend shows the values of applied pre-shear rates ($\dot{\gamma}_{PS}$) before the flow cessation.

After establishing a steady homogeneous shearing flow, we then impose a sudden step-down jump in the shear rate to arrest the flow. However, owing to instrument inertia, it takes a finite time for the instantaneous shear rate ($\dot{\gamma}(t)$) to approach zero. In Fig. 9, we plot the damped inertial decay of the instantaneous shear rate for both materials as a function of time since the cessation of flow. The inserts show the same data on greatly expanded ordinate scales, and reveal that after approximately 0.1 s, only small random fluctuations in the instantaneous shear rate remain. Subsequently, we monitor the evolution of the shear stress exerted by the sample for an hour, and this data is plotted in Figs. 10(a) and 10(b), respectively for the clay dispersion and silica gel. It can be seen that in both systems, irrespective of the prior steady-state pre-shear rate, there is a non-monotonic evolution in the shear stress with time. However, while there are some similar characteristics (an initial decrease in the stress followed by a subsequent increase), the specific behavior in each sample is quite distinct.

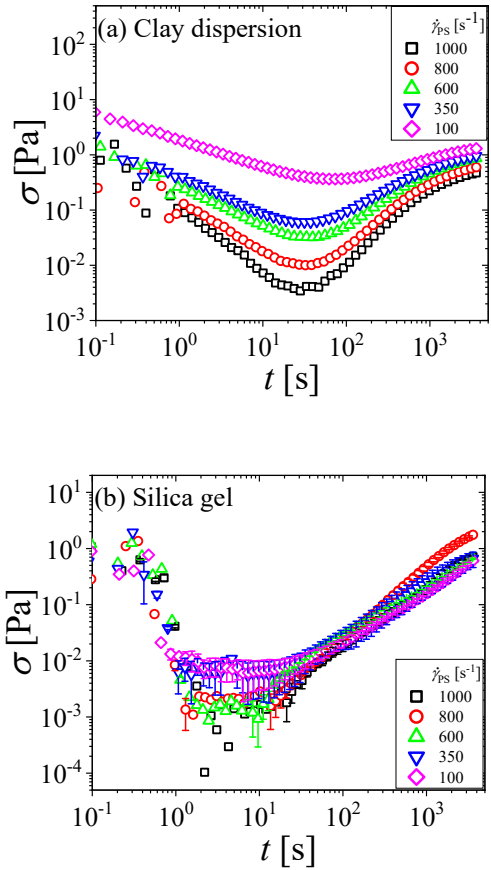

Figure 10: Evolution of shear stress after the cessation of steady shearing flow for (a) the clay dispersion and (b) the silica gel. The legends show the values of the applied pre-shear rate ($\dot{\gamma}_{PS}$) before the flow cessation.

Interestingly, the shear stress evolution in the clay dispersion strongly depends on the pre-shear rate ($\dot{\gamma}_{PS}$), as shown in Fig. 10(a). It can be seen that the initial decrease or *relaxation* of the shear stress gets faster with an increase in the pre-shear rate. Subsequently, the shear stress evolution curves associated with all values of the pre-shear rate show a minimum around the same time (≈ 40 s) and then show an increasing trend, consistent with a shear stress build-up. Remarkably, the value of the shear stress value associated with this minimum decreases, whereas the rate of increase in the shear stress subsequent to the minimum point is observed to increase with an increase in pre-shear rate. Therefore, the dynamical behavior of the clay dispersion suggests that the intensity of the shear stress build-up phenomenon diminishes with a decrease in the pre-shear rate.

Contrary to what observed for clay dispersion, the effect of the shear rate during pre-shear has a much weaker effect on the silica gel, as shown in Fig. 10(b). For the silica gel, after the cessation of flow, the shear stress undergoes a decrease, shows a minimum at around 3 – 5 s, and subsequently undergoes an increase following broadly the same path for all values of the pre-shear rate. The minimum value of the shear stress for the silica gel is in the range of $10^{-2} - 10^{-3}$ Pa (very close to the limits of rheometer sensitivity for this geometry) and seems to increase (weakly) with a decrease in pre-shear rate. After passing through the minimum, beyond 30 s all the shear stress build-up curves can be seen to follow the same path irrespective of the magnitude of the pre-shear rate.

In addition to monitoring the shear stress evolution by imposing a shear rate of $\dot{\gamma} = 0$ s$^{-1}$, we also subject both the materials to a small strain amplitude oscillatory flow field in the linear domain at $\omega \in [0.5 - 10]$ rad/s subsequent to cessation of steady state pre-shearing over the range of pre-shear rates. Fig. 1 suggests that for the clay dispersion, the maximum initial increase in elastic modulus ($G'$) is observed at low frequency $\omega = 0.5$ rad/s, whereas the rate of evolution of $G'$ for the silica gel is largely independent of the imposed oscillatory frequency $\omega$. For consistency and clarity the corresponding time evolution of the age- and history-dependent storage modulus $G'(t; \omega, \dot{\gamma}_{PS})$ is plotted in Fig. 11 for a constant angular frequency 0.5 rad/s and for different pre-shear rate for both materials (for other angular frequencies, refer to supplementary information Figs. S4 and S5). In general, for both the materials, this history-dependent storage modulus increases with elapsed time since the pre-shearing ceased. It can be seen that for the clay dispersion, with an increase in pre-shear rate the initial value of $G'$ begins at a lower value. However, as time elapses, the value of the instantaneous elastic modulus associated with higher pre-shear rates increases at a faster rate and beyond 1000 s, all of the evolving curves for $G'(t; \omega, \dot{\gamma}_{PS})$ associated with different pre-shear rate merge.

For the silica gel, on the other hand, the evolution of $G'$ is observed to be practically identical for all the explored values of pre-shear rate. In the inset of Fig. 11 (a) and 11 (b), the logarithmic rate of microstructural aging (LROMA) corresponding to the slope $(\partial \ln G' / \partial \ln t)_\omega$ with respect to time is plotted for different pre-shear rates. We evaluate the results for both the clay dispersion and the silica gel over two different time intervals; short times (determined from data spanning 80 – 200 s) and at later times when the samples are older (2000 – 3600 s). From these inset figures, it can be seen that for clay dispersion the initial rate of evolution of $G'$ shows a strong increase with pre-shear rate. However, at later times following flow cessation, the rate of evolution decreases significantly and is only a weak function of pre-shear rate. Interestingly, for the silica gel, while the rate of evolution of $G'$ decreases with time, it remains independent of the pre-shear rate. Overall the behavior suggests more rapid structural recovery for higher values of pre-shear rate. For silica gel, microstructure recovery is practically unaffected by pre-shear rate, resulting in an almost identical evolution of $G'$.

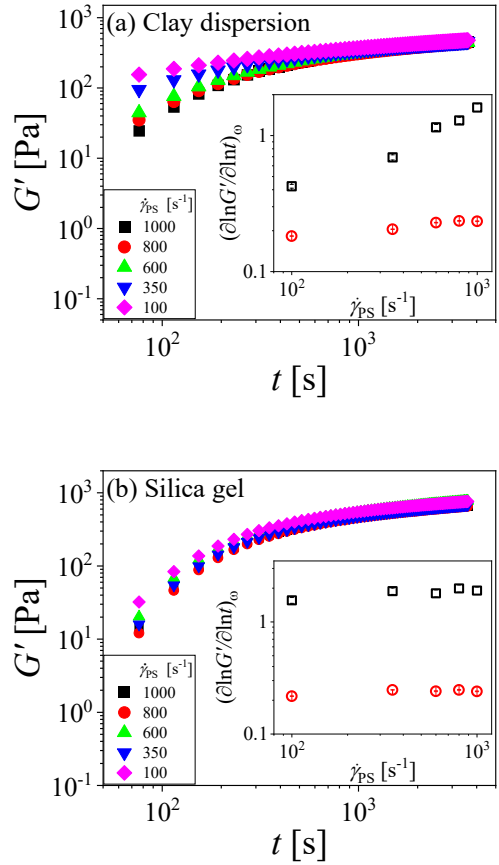

Figure 11: Evolution of the age- and history-dependent elastic storage moduli ($G'$) with time at $\omega = 0.5$ rad/s after flow cessation from different pre-shear rates for (a) the clay dispersion and (b) the silica gel. The legend in each figures shows the different pre-shear rates imposed before the flow cessation. The inset figures in (a) and (b) show the variation in the logarithmic rate of microstructural aging (($\partial \ln G' / \partial \ln t)_\omega$) (LROMA) with respect to pre-shear rate over two different time domains (black open squares (80 – 200 s) and red open circles (2000 – 3600 s)) for the clay dispersion and silica gel respectively.

## IV. Discussion

The results discussed in the previous sections can be divided into two parts. The first part provides a thorough characterization of both materials, which leads to the identification of specific soft glassy rheological attributes associated with the aqueous clay and colloidal silica dispersions. The second part identifies some of the characteristic features of the shear stress build-up behavior observed in both systems following cessation of a strong letherizing or shear-melting deformation history. The key features of the soft glassy rheological behaviour of the clay dispersion and silica gel can be summarised as follows:

1. After cessation of shear melting, the history-dependent elastic modulus ($G'$) increases as a function of time as shown in Fig. 1. Initially, it shows a strong increase followed by a weak increase, and both behaviors seem to follow a power law dependence. In the clay dispersion, for a given pre-shear rate, the subsequent rate of increase of $G'$ with elapsed time (since the cessation of pre-shearing) is faster at lower frequencies $\omega$ (Fig. 1 (a)); whereas for a fixed $\omega$, the evolution of $G'$ with time becomes weaker with decreases in the imposed pre-shear rate (Fig. 11 (a)). Interestingly, for the silica gel, the history-dependent evolution of $G'$ with time remains independent of the oscillatory probe frequency $\omega$ (Fig. 1(b)) and imposed pre-shear rate (Fig. 11(b)) over the ranges tested.
2. The yield stress remains initially constant for a period of time after cessation of the shear melting, and then subsequently increases with time (Fig. 4). Viscosity bifurcation experiments show that the corresponding yield stress or critical flow stress determined from this deformation history is of similar magnitude. The corresponding steady-state shear rate determined from the lowest value of the imposed yield stress (or minimum flow stress) provides a useful measure of the minimum value of the shear rate for which a homogeneous flow can be established in each system (Fig. 5).
3. The creep curves measured at different waiting times show time – waiting-time superposition (TWTS) in a suitably-transformed effective time domain, and the corresponding master curves are consistent with a power-law dependence of the characteristic viscoelastic relaxation time on the age time for both materials. The corresponding power-law coefficients are $1.89 \pm 0.05$ and, $1.24 \pm 0.06$ respectively, suggesting both materials show hyper-aging behavior (Fig. 8).

Structural kinetic models that incorporate viscoelasticity and physical aging along with an age-dependent increase in the elastic modulus have been observed to predict shear stress build-up similar to the features we have documented here [50, 57]. However, in this study, rather than down selecting to a specific comprehensive microstructurally-based description and the associated definition of a suitable structural kinetic model (as has been done before), we adopt a simple data-driven approach that leads to an analytical solution characterizing the observed behavior. In this empirically-based methodology, we thus consider a simple linear Maxwell model with time-dependent modulus and relaxation time (and thus also a time evolving viscosity $\eta(t) = G(t)\tau(t)$). The corresponding viscoelastic constitutive equation is thus given by:

$$\dot{\gamma} = \frac{d}{dt}\left(\frac{\sigma}{G(t)}\right) + \frac{\sigma}{G(t)\tau(t)}. \tag{5}$$

The age-dependent modulus is assumed to show a power-law dependence on the sample age and is given by:

$$G(t) = G_0(t/\tau_m)^q, \tag{6}$$

where $G_0$ is the elastic modulus at a time $\tau_m$ after shear melting has ceased, and $q$ is the power law exponent that characterizes the logarithmic rate of microstructural aging (i.e. the exponent we identified experimentally from the function LROMA) associated with the model. We consider the time-dependent relaxation time associated with the underlying linear viscoelastic Maxwell model to be the same as given by Eq. 3.

$$\tau = A_1 \tau_m (t/\tau_m)^\mu.$$

Under the application of constant shear rate $\dot{\gamma} = \dot{\gamma}_{PS}$ during shear melting, the material acquires a steady-state shear stress that depends on the pre-shear rate. We stop the shear melting at a certain time by applying $\dot{\gamma} = 0$ and we reinitialize the time ($t = 0$) at this instant. Eqs. 5, 6, and 3, then can be solved for shear stress leading to:

$$\sigma(t) = \sigma_1 \left(\frac{t}{\tau_m}\right)^q \exp\left[\frac{1}{A_1(\mu-1)}\left\{\frac{1}{\left(\frac{t}{\tau_m}\right)^{\mu-1}} - 1\right\}\right]. \tag{7}$$

Where $\sigma = \sigma_1$ is the shear stress at $t = \tau_m$ that depends on the pre-shear rate ($\dot{\gamma}_{PS}$). It has been proposed that immediately after the shear rate is set to zero, it takes a time of the order of microscopic timescale to readjust to the new state such that Eqs. 3 and 6 become applicable for $t \geq \tau_m$. The microscopic timescale $\tau_m$ is proposed to follow an Arrhenius dependence on temperature, expressed as: $\tau_m = \tau_{m0}\exp(U/k_B T)$, where $\tau_{m0}$ denotes an attempt time, and $U$ corresponds to the energy barrier associated with the microscopic motion of the entity within the cage [67, 81]. Typically, the attempt time $\tau_{m0}$ is expected to be around $10^{-1}$ s. By analyzing the evolution of relaxation time and elastic modulus at different temperatures, Joshi estimated $U$ for 54 days old aqueous Laponite® dispersion having 2.8 wt. % concentration and reported it to be around $25\, k_B T$, where $T = 298.15\, K$ (The behavior of this system is expected to be closer to the present system of 30 days dispersion having 3.5 wt.% concentration). The corresponding $\tau_m$ comes out to be $0.0072\, s$ [67]. For Ludox® gel, Kumar and Joshi estimated $U$ for the same system used in this work to be around $30\, k_B T$, which leads to $\tau_m$ to be around 1 s [82].

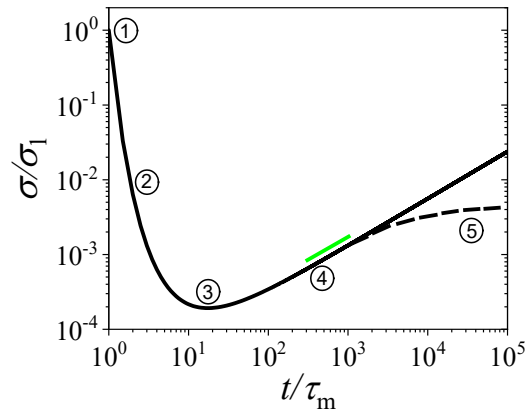

Figure 12: The evolution of the normalized shear stress ($\sigma/\sigma_1$) after the cessation of steady shear rate is plotted against the normalized time ($t/\tau_m$) for the time-dependent Maxwell model. Refer to the text for more details including numbers marked on the plot. The model parameter values used for the prediction are: $q = 0.63, \mu = 2$, and $A_1 = 0.091$.

In Fig. 12 we show the solution to the equation in terms of a dimensionless stress $\sigma/\sigma_1$ as a function of $t/\tau_m$ (scaled with the reference scale $\tau_m$). It can be seen that the model qualitatively matches the experimentally observed behavior shown in Fig. 10 very well. Various aspects of the shear stress behavior are marked on the figure by numbers. The analysis of these aspects is as follows:

1. The initial value of the shear stress, from which the solution starts to decay, is associated with the (monotonic) steady-state flow curve $\sigma_{ss}(\dot{\gamma})$ and is controlled by the steady-state shear rate $\dot{\gamma}_{PS}$ applied during the pre-shear step. The value $\sigma_1$ is stress in the sample after $\tau_m$ time has elapsed. Nonetheless, the greater the shear rate during the pre-shear step, the greater the initial shear stress $\sigma_1$ will be.

2. At short times after the pre-shear is removed, the relaxation of the shear stress is associated with the exponential term in equation (7). The precise functional form thus depends on the value and evolution of the relaxation time $\tau(t)$ over the explored time. Consequently, it strongly depends on the values of $A_1$ and $\mu$.

3. The minimum associated with the shear stress relaxation curve can be obtained by taking the derivative of $\sigma/\sigma_1$ with respect to $t/\tau_m$ in equation (7) and equating it to zero. This analysis leads to the value of $t/\tau_m$ at the shear stress minimum being determined as:

$$\left[\frac{t}{\tau_m}\right]_{min} = (A_1 q)^{1/(1-\mu)} \tag{8}$$

Since $\mu > 1$ for a hyperaging material, this expression clearly indicates that the greater the values of $q$, and $A_1$, the smaller the time at which the minimum occurs.

4. This region is associated with the build-up part of the shear stress, and from equation (7), it is clearly connected to the power law term $(t/\tau_m)^q$. From the tangent line drawn in Fig. 12 in region (4) we can thus empirically define a *logarithmic rate of stress evolution* (LROSE) as $\partial \ln \sigma / \partial \ln t$. Interestingly in our model, the shear stress build up measured at long times following the cessation of shear flow is predicted to grow in time with the same power law coefficient ($q$) as that observed for the linear viscoelastic modulus. (i.e. for the model considered herein we expect LROSE = LROMA). However, this may not be true for all materials and can be verified experimentally. For those materials that do not show time dependence of the elastic modulus ($q = 0$), Eq. (7) will not predict any shear stress build-up, but instead predict eventual complete shear stress relaxation for ($\mu < 1$) and logarithmic relaxation for $\mu = 1$. The scenario of hyperaging dynamics ($\mu > 1$) is intriguing as the relaxation time scale characterizing the evolution of the shear stress grows stronger than linearly with respect to waiting time. This makes complete relaxation of a material impossible and provides a pathway for generating permanent residual stresses in the sample [50, 83].

5. The dashed line shown in the figure is a schematic representation of what may happen if the rate of growth of modulus in time continues to decrease. The experimental data, particularly for the clay dispersion, shows very similar behavior.

For real materials, the simple power-law age-time dependence of elastic modulus expressed by Eq. (6) may not be applicable for the entire duration of such an experiment. As shown in in Figs. 1 and 11, the elastic storage modulus $G'(t; \omega)$ (i.e., the logarithmic rate of microstructural aging, LROMA) initially shows a rapid increase followed by a weaker increase at larger waiting times. More specifically, for the clay dispersion, the greater the magnitude of the initial pre-shear rate, the stronger the initial increase in the dynamic storage modulus is (Fig. 11 (a)). For the silica gel, the change in elastic modulus with waiting time is not affected by pre-shear rate (Fig. 11 (b)).

We also note that both systems follow the same dependence for relaxation time given by Eq. (3). However, this dependence applies only beyond a waiting time of 100 s (Figs. 7 and 8). For shorter waiting times, the viscoelastic relaxation time may grow more strongly (i.e. with larger values of $\mu$ or possibly even an exponential dependence [84]) than what has been observed at later times. This dependence may also vary with the pre-shear rate; however, there is not an easy analytic way to incorporate such complications in this simple illustrative model. For more realistic modelling, different functional dependencies could be considered for different waiting time durations, and the resulting coupled differential-algebraic equations can be integrated numerically. However, in the present work, for simplicity we assume Eqs. (6) and (3) are uniformly valid expressions for the elastic modulus and the relaxation time. For the elastic modulus, we consider different values of $G_0$ and $q$ for different shear melting conditions, as mentioned in Fig. 11.

The stress evolution predicted, shown by Eq. (7), suggests that the rate at which the shear stress increases is directly related to the rate at which modulus increases. Fig. 11 shows that the elastic modulus undergoes an initial rapid increase followed by a slow increase . For the clay dispersion, the rate at which the elastic modulus initially increases strongly depends on the pre-shear rate and the angular frequency used during the aging experiment. However, the rate of increase in the elastic modulus during the slow increase domain does not depend on the shear rate applied during the pre-shear and also becomes independent of the frequency. For the silica gel, the evolution in the elastic modulus in both regimes is independent of the shear rate applied during the pre-shear as well as the frequency. Furthermore, the two regimes of elastic modulus increase are distinct over a time window of 80 to 200 s and 2000 to 3600 s. Interestingly, as discussed before, the shear stress build-up behavior shown in Fig. 10 for clay dispersion depends on the shear rate applied during the pre-shear, while for silica gel, the stress build-up behavior is independent of the shear rate applied during the pre-shear.

To summarize these complex dependencies, in Fig. 13 we plot the experimentally-determined logarithmic rate of stress evolution (LROSE), i.e. $(\partial \ln\sigma / \partial \ln t)$ as a function of the experimentally-determined logarithmic rate of microstructural aging (LROMA) i.e. $(\partial \ln G' / \partial \ln t)_\omega$ in two different time regions for both systems at different pre-shear rates.

Since, in the clay dispersion, the rate of evolution of $G'(t; \omega)$ with time also shows an inherent frequency dependence, we plot the logarithmic rate of microstructural aging for both $= 0.5$ rad/s, and $\omega = 10$ rad/s, which bound the maximum and minimum values over the explored frequency range. Whereas, for silica gel, the logarithmic rate of microstructural aging is plotted for only $= 0.5$ rad/s, as the evolution of the elastic storage modulus remains independent of angular frequency. For the clay dispersion, the five points associated with each time zone represent the five different shear rates applied during the pre-shear. Interestingly, all the points in both time regions show that the logarithmic rate of stress evolution (LROSE) during the observed shear stress build-up is on the same order of magnitude as the logarithmic rate of microstructural evolution (LROMA) measured using SAOS over the same time region.

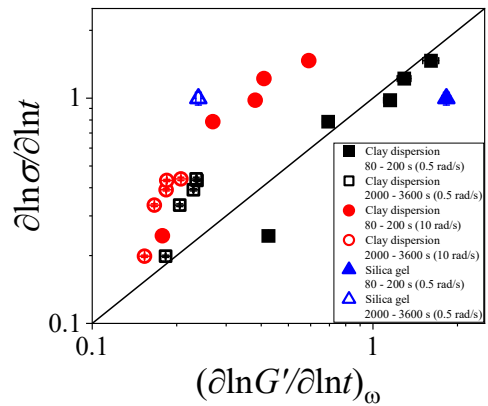

Figure 13: The logarithmic rate of stress evolution $(\partial \ln\sigma / \partial \ln t)$ (LROSE) after the flow cessation from different pre-shear rates in the range of $100 \leq \dot{\gamma}_{PS} \leq 1000$ s$^{-1}$ for clay dispersion and silica gel is plotted against the logarithmic rate of microstructural aging (LROMA) measured using small amplitude oscillatory deformations of sample aging $((\partial \ln G'(t; \omega) / \partial \ln t)_\omega)$. The different symbols in the legend identify the system, angular test frequency ($\omega$), and the sample age time over which these parameters were obtained.

For the silica gel, within experimental uncertainty, the shear stress build-up (LROSE) occurs at an identical rate for all shear rates applied during the pre-shear throughout the entire explored range, as shown in Fig. 10. However, the elastic modulus does increase at two different (logarithmic) rates at early and later times (as indicated by the filled and hollow triangles in Fig. 11). Interestingly, however, the LROMA is independent of the shear rates applied during the pre-shear and the angular frequencies used during the aging experiment.

From the data representation in Fig. 11 it is clear that in both systems, there appears to be a direct relationship between the evolution of the dynamic elastic modulus measured in aging experiments and the residual shear stress measured following the cessation of strong pre-shear. However, the dependence is more complex than the simple linear parity line predicted by the simple aging Maxwell model (Eq. (7)). Even for the (somewhat simpler) rate-independent silica gel sample, the logarithmic rate of stress growth (LROSE) is only observed to be of the same order of magnitude as the value of LROMA at short times (filled triangles). Interestingly, the value of $\partial \ln \sigma / \partial \ln t$ does not appear to change much at longer times, even though the value of LROMA decreases with sample age. This representation in terms of independently observable logarithmic rates of stress evolution and microstructural aging may be useful in the future for comparing and contrasting different thixotropic elastoviscoplastic constitutive models.

Interestingly many soft glassy materials have been reported to demonstrate an intriguing rheological phenomenon termed *overaging* [19, 44, 85, 86]. The origin of this behavior lies in the modification of the relaxation time spectra, wherein the application of moderate magnitudes of oscillatory strain alters the relaxation time spectra in such a fashion that the number of short relaxation time modes decreases while slow relaxation time modes become increasingly populated, leading to an enhancement of the viscosity. The Soft Glassy Rheology model has been reported to predict overaging behavior very well [19, 44, 85, 86]. However, the phenomenon of stress build-up should not be confused with overaging which we believe from our experimental measurements and simple model to originate primarily from the age-dependent increase in the elastic modulus and the hyper-aging ($\mu > 1$) dynamics of soft glassy materials.

The discussion based on the time-dependent single-mode Maxwell model suggests that the origin of shear stress buildup lies in the enhancement of the modulus during physical aging. Our experiments with clay dispersion further showed that flow cessation (or *quenching*) from a larger value of the pre-shear rate leads subsequently to a more rapid increase in the elastic modulus. This occurs because a greater magnitude of the pre-shear causes a more significant breakdown of the material microstructure (and larger entropic disorder), resulting in more rapid reformation and a stronger enhancement of the elastic modulus. This is confirmed by comparing two derivatives: the LROSE ($\partial \ln\sigma / \partial \ln t$) and LROMA ($\partial \ln G' / \partial \ln t)_\omega$. For clay dispersion, LROMA varies by an order of magnitude across frequencies, while for silica gel, it remains nearly constant.

The 1-to-1 correspondence of the values of LROSE and LROMA comes from our use of a single-mode time-dependent Maxwell model. In practical situations, the stress buildup phenomena will be influenced by how the modulus values associated with different relaxation modes (or how the weights of a continuous relaxation spectrum $H(\tau, t_w)$) evolve upon flow cessation and how the stress buildup is affected by the same. This requires consideration of a more complete structural kinetic formalism involving a spectrum of relaxation and thixotropic modes, which we plan to address in future work. Additionally, we will compare the stress buildup observed in the present work with what has been reported in the literature. Interestingly, two earlier works [46, 49] suggest that the intensity of stress buildup increases with a decrease in the magnitude of the shear rate during pre-shear. Conversely, our present work for two different colloidal materials, as well as Negi and Osuji's [53] work, suggests that greater pre-shear intensity leads to a more pronounced stress buildup. The present work also indicates that the presence of a non-homogeneous flow field (shear banding) during the pre-shear step may not be a prerequisite for stress buildup. The empirical observations of the present work are also corroborated by the proposed single-mode time-dependent Maxwell model. However, the conditions under which this dependence is reversed, as observed in the previous two studies, need to be explored, and we also leave this task for future work.

## V. Conclusions

In this work, we have studied the aging dynamics of two soft glassy materials: a 30 days old 3.5 wt % aqueous Laponite® RD dispersion (clay dispersion) and a 2 day old aqueous Ludox® gel (silica gel). We focus on the evolution of the rheological behavior with sample age and shear rate, particularly the peculiar shear stress buildup these materials show subsequent to the cessation of steady shear flow. A thorough rheological characterization of these materials reveals that upon cessation of steady shear flow, the elastic modulus of both materials increases in two stages: an initial rapid growth followed by a slower, prolonged increase at longer times. Interestingly, the intensity of this initial rapid growth in the elastic modulus increases with the shear rate applied during pre-shear for the clay dispersion. For the silica gel, however, the aging of the elastic storage modulus remains independent of the shear rate applied during pre-shear. We also measured the static yield stress of both materials as a function of the waiting time elapsed since the cessation of steady shear flow. The static yield stress remains constant for a brief period after the flow stops, beyond which it continuously increases.

Creep experiments performed at different waiting times result in a creep – time-waiting time superposition, consistent with a power law dependence of the characteristic relaxation time (or the average value of an underlying relaxation time spectrum) on the age time. We observe that the power law exponent is greater than unity for both materials, suggesting hyper-aging dynamics. All of these characteristic features, namely the time dependence of the elastic modulus and the static yield stress as well as the hyper-aging dynamics of the relaxation time, implicitly indicate that the steady-state shear stress-shear rate flow curves of both materials are non-monotonic, demonstrating a shear stress minimum. Both materials also exhibit a viscosity bifurcation in creep experiments; below a critical shear stress, the instantaneous viscosity $\eta^+ = \sigma_0/\dot{\gamma}(t - t_w)$ continuously increases, leading to the eventual cessation of flow. The minimum observable steady shear rate, $\dot{\gamma}_c = \min\{\dot{\gamma}_{ss}\}$ and the corresponding critical shear stress $\sigma_c$ consequently identify the locus of the minimum in the steady-state flow curve given by $(\dot{\gamma}_{ss})$.

Finally, we conducted flow cessation or *quench* experiments, monitoring the evolution of shear stress following the stoppage of steady shear flow. The shear rates used during the pre-shear belonged to the monotonic part of the flow curve to avoid complications of steady-state shear banding. Interestingly, we observed that both colloidal materials (clay dispersion, and silica gel), exhibited an initial shear stress relaxation (a decrease in shear stress with time) followed by a shear stress build-up, resulting in a distinct local minimum in the shear stress. For the clay dispersion, the rates of both the initial shear stress relaxation and the subsequent shear stress build-up increased with the magnitude of the shear rate applied during pre-shear. Consequently, the minimum in the shear stress evolution curve shifted to higher stress values with a reduction in the pre-shear rate. In contrast, for the silica gel, the initial shear stress relaxation, which was much more rapid compared to the clay dispersion, and the subsequent shear stress build-up were observed to be independent of the shear rate applied during pre-shear.

We find empirically that the logarithmic rate of shear stress build-up following the strong quench is of the same order of magnitude as the logarithmic rate of aging in the microstructure (as determined by measurements of the elastic modulus using small amplitude oscillatory shear). To help understand the underlying dynamics, we propose a simple time-dependent linear Maxwell model in which both the elastic modulus and the relaxation time are considered to follow the independently observed dependences on sample age. This model captures many of the features observed in the quench experiments, including a rapid initial viscoelastic decay in the stress, a local minimum, and then subsequent stress growth that can result in permanent residual stress in a strongly pre-sheared sample. Overall, this study comprehensively investigates the rheological behavior of two model soft glassy materials and helps rationalize the puzzling shear stress buildup behavior observed experimentally. The double-logarithmic representation of the measured dynamics in rapid quenches and in conventional viscoelastic aging tests may be helpful in helping to distinguish the predictions of different thixotropic elastoviscoplastic (TEVP) constitutive models.

## Acknowledgement:

YMJ would like to acknowledge financial support from the Science and Engineering Research Board, Government of India (Grant Nos. CRG/2022/004868 and JCB/2022/000040). GHM acknowledges financial support from the MIT School of Engineering.

## Author Declarations:

The authors have no conflicts to disclose.

# Supplementary information

# Shear Stress Build-up Under Constant Strain Conditions in Soft Glassy Materials

Vivek Kumar[1], Gareth H McKinley[2], and Yogesh M Joshi[1,*]

[1]Department of Chemical Engineering, Indian Institute of Technology Kanpur, Kanpur, Uttar Pradesh, India, 208016.

[2]Hatsopoulos Microfluids Laboratory, Department of Mechanical Engineering, Massachusetts Institute of Technology, Cambridge, Massachusetts, USA, 02139

[*]Corresponding Author, email: joshi@iitk.ac.in

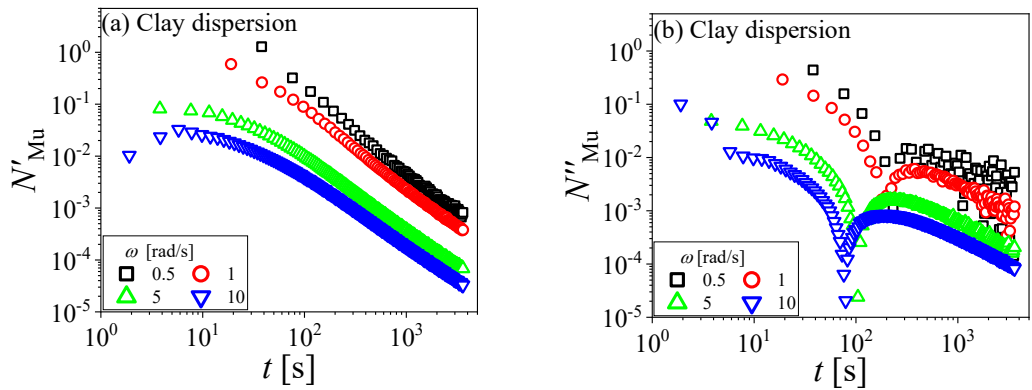

Figure S1: Evolution of (a) Elastic ($N'_{\text{Mu}}$) and (b) viscous ($N''_{\text{Mu}}$) mutation number with time ($t$) at different angular frequencies ($\omega$) (values are given in the legend) for the clay dispersion after pre-shearing the system at 1000 s$^{-1}$.

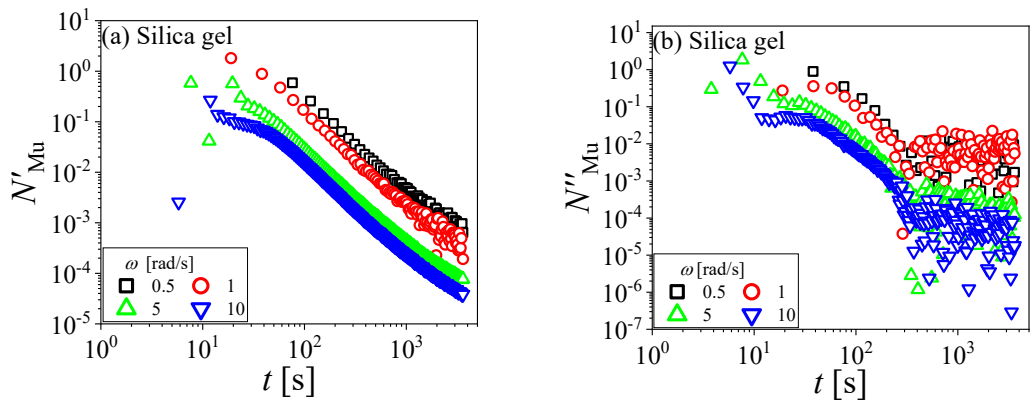

Figure S2: Evolution of (a) Elastic ($N'_{\text{Mu}}$) and (b) viscous ($N''_{\text{Mu}}$) mutation number with time ($t$) at different angular frequencies ($\omega$) (values are given in the legend) for the silica gel after pre-shearing the system at 1000 s$^{-1}$.

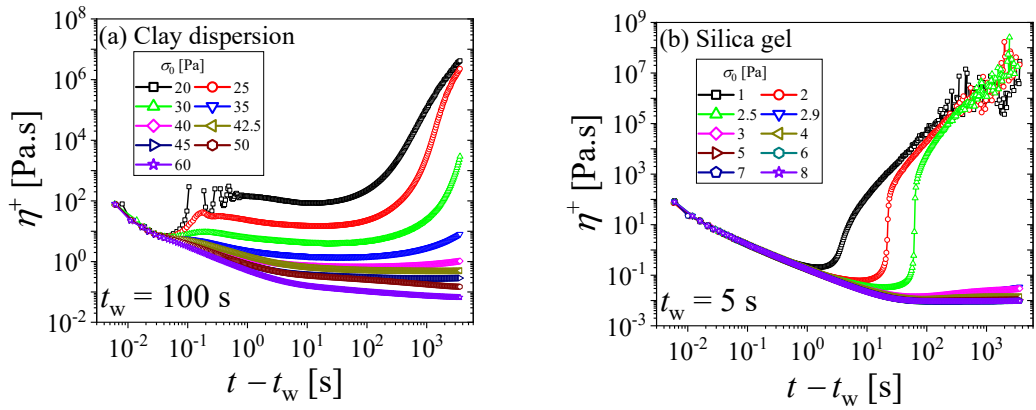

Figure S3: Evolution of apparent viscosity ($\eta^+ = \sigma_0/\dot{\gamma}(t - t_\mathrm{w})$) with time at different applied shear stresses (values are shown in the legend) for (a) clay dispersion and (b) silica gel. During the waiting time ($t_\mathrm{w}$), an oscillatory shearing field with 0.1% strain amplitude and 0.63 rad/s angular frequency was applied.

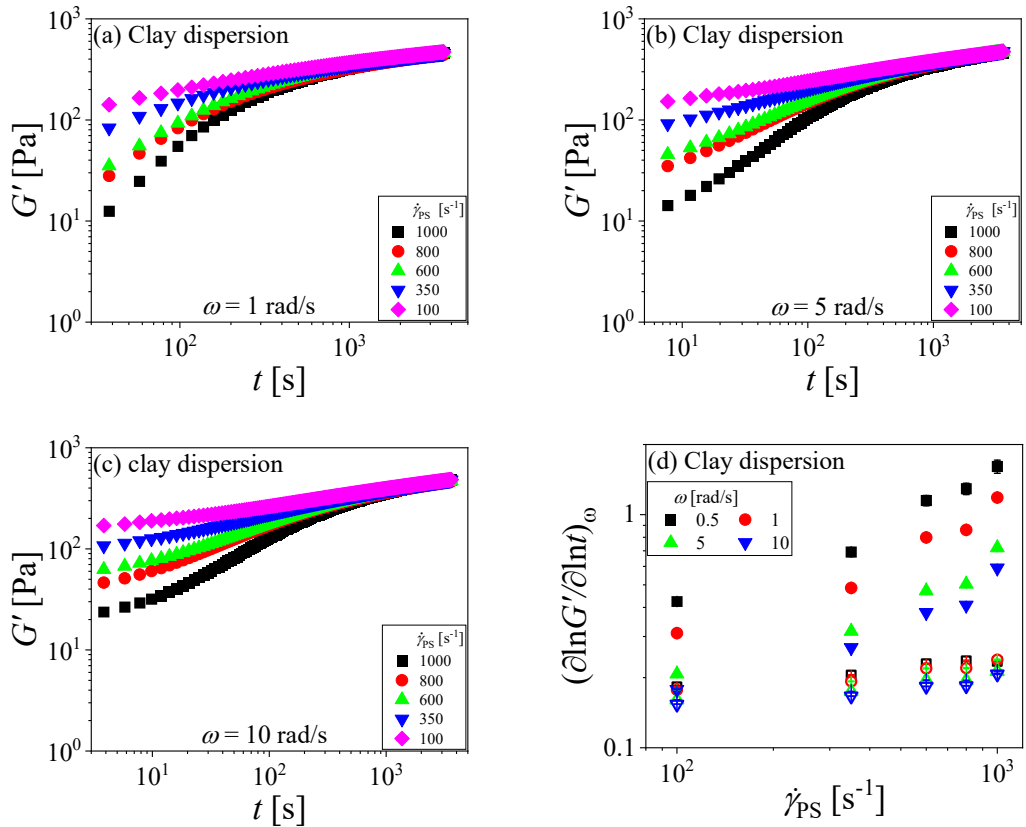

Figure S4: Evolution of age- and history-dependent elastic storage moduli ($G'$) with time after flow cessation from different pre-shear rates ($\dot{\gamma}_{PS}$) (values are shown in the legend) is plotted for clay dispersion at (a) = 1 rad/s, (b) $\omega$ = 5 rad/s and (c) $\omega$ = 10 rad/s. In (d), the logarithmic rate of microstructural aging ($(\partial \ln G'/\partial \ln t)_\omega$) values are plotted against the pre-shear rate over two different time domains 80 – 200 s (closed symbols) and 2000 – 3600 s (open symbols) for different angular frequencies ($\omega$) (values are shown in the legend).

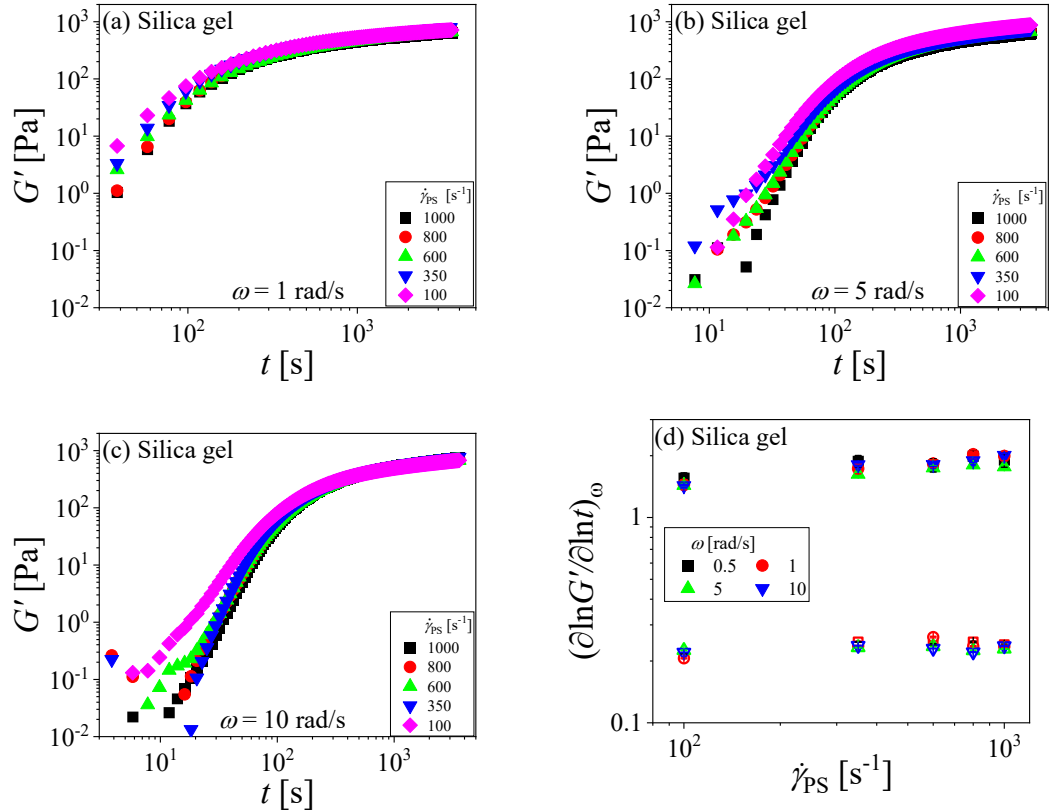

Figure S5: Evolution of age- and history-dependent elastic storage moduli ($G'$) with time after flow cessation from different pre-shear rates ($\dot{\gamma}_{PS}$) (values are shown in the legend) is plotted for silica gel at (a) = 1 rad/s, (b) $\omega$ = 5 rad/s and (c) $\omega$ = 10 rad/s. In (d), the logarithmic rate of microstructural aging (($\partial \ln G'/ \partial \ln t)_\omega$) values are plotted against the pre-shear rate over two different time domains 80 – 200 s (closed symbols) and 2000 – 3600 s (open symbols) for different angular frequencies ($\omega$) (values are shown in the legend).